\documentclass[article, twocolumn,
   aps, pra,
  amsmath,amssymb,
  longbibliography,
  ]{revtex4-1}
\usepackage{graphicx,color}
\usepackage{amsmath}
\usepackage{natbib}
\usepackage{epsfig}
\begin{document}

\title{Note: Stokes-Einstein relation without hydrodynamic diameter in the TIP4P/Ice water model}

\author{S. A. Khrapak}\email{Sergey.Khrapak@gmx.de}
\affiliation{Joint Institute for High Temperatures, Russian Academy of Sciences, 125412 Moscow, Russia}
\author{A. G. Khrapak}
\affiliation{Joint Institute for High Temperatures, Russian Academy of Sciences, 125412 Moscow, Russia}

\begin{abstract}
It is demonstrated that self-diffusion and shear viscosity data for the TIP4P/Ice water model reported recently [L. Baran, W. Rzysko and L. MacDowell, J. Chem. Phys. {\bf 158}, 064503 (2023)] obey the microscopic version of the Stokes-Einstein relation without the hydrodynamic diameter.  
\end{abstract}

\date{\today}

\maketitle

The conventional macroscopic Stokes-Einstein (SE) relation connects the diffusion coefficient $D$ of a tracer macroscopic spherical (``Brownian'') particle of radius $R$ with the temperature $T$ and the shear viscosity  coefficient $\eta$ of a fluid it is immersed in~\cite{BalucaniBook}:
\begin{equation}\label{SE_1}
D=\frac{k_{\rm B}T}{6\pi \eta R}.
\end{equation}   
Here $k_{\rm B}$ is the Boltzmann' constant and the numerical coefficient in the denominator corresponds to the ``stick'' boundary condition at the sphere surface. When the size of the sphere decreases and becomes comparable to the characteristic distance of intermolecular interactions, it is common to replace the actual size of the sphere by the so-called hydrodynamic radius $R_{\rm H}$, which is often considered as an adjustable parameter. On the microscopic level, when self-diffusion of atoms in simple pure fluids is considered, the SE relation becomes
\begin{equation}\label{SE_2}
D\eta\left(\frac{\Delta}{k_{\rm B}T}\right)=\alpha_{\rm SE},
\end{equation}
where $\Delta=\rho^{-1/3}$ is the mean interatomic separation and $\rho$ is the atomic number density. It is easy to see that the role of the effective tracer sphere diameter is now played by the characteristic interatomic separation $\Delta$. Equation (\ref{SE_2}) is sometimes referred to as the SE relation without the hydrodynamic diameter~\cite{CostigliolaJCP2019}. We adopt this terminology here. The numerical coefficient $\alpha_{\rm SE}$ is only weakly system and state-dependent, as discussed in more detail below.

A relation of the form of Eq.~(\ref{SE_2}) was already discussed by Frenkel~\cite{FrenkelBook} in connection with the viscosity of simple liquids. He provided some qualitative arguments regarding why a macroscopic approach might be applied down to the atomic scale. Quantitative models, explaining why relation of the form of Eq.~(\ref{SE_2}) should work in simple fluids have been put forward later~\cite{ZwanzigJCP1983,Balucani1990}. In particular, Zwanzig's model provides an explicit expression for the SE constant: $\alpha_{\rm SE}\simeq 0.132(1+\eta/2\eta_{l})$, where $\eta_l$ is the longitudinal viscosity. This implies that $\alpha_{\rm SE}$ can vary
only between $0.132$ and $0.181$. Perhaps, more insight can gained by rewriting the SE coefficient as $\alpha_{\rm SE}\simeq 0.132(1+c_t^2/2c_l^2)$, where $c_t$ and $c_l$ are the high-frequency transverse and longitudinal sound velocities, respectively~\cite{KhrapakMolPhys2019}. In  very soft plasma-related systems with Coulomb-like interactions the inequality $c_t\ll c_l$ holds and $\alpha_{\rm SE}$ should tend to its lower limit. For hard-sphere like interactions the velocity ratio approaches $c_t/c_l\sim 0.5$~\cite{KhrapakPRE09_2019,KhrapakPRE05_2021} and $\alpha_{\rm SE}$ should  increase towards the upper limit. This correlation between the magnitude of $\alpha_{\rm SE}$ and the softness of the interaction potential is generally rather well reproduced~\cite{KhrapakMolPhys2019,KhrapakPRE10_2021}. Remarkably, no concepts of hydrodynamic diameter or boundary condition are required in this microscopic formulation of the SE relation.         

Over the years, numerous confirmations of the applicability of the microscopic SE relation without the hydrodynamic diameter to dense simple fluids have been reported. This concerns single component Coulomb (one-component plasma) and screened Coulomb (complex or dusty plasma) fluids of charged particles~\cite{DaligaultPRL2006,DaligaultPRE2014,KhrapakAIPAdv2018,
KhrapakMolecules12_2021}, soft (inverse power) repulsive particle fluid~\cite{HeyesPCCP2007}, Lennard-Jones fluid~\cite{CostigliolaJCP2019,OhtoriPRE2015,OhtoriPRE2017}, Weeks-Chandler-Andersen fluid~\cite{OhtoriJCP2018}, and the hard sphere fluid~\cite{OhtoriJCP2018,Pieprzyk2019,KhrapakPRE10_2021}. A few recent examples confirming applicability of SE relation in the form of Eq.~(\ref{SE_2}) to real liquid substances include liquid iron at
conditions of planetary cores~\cite{LiJCP2021}, dense supercritical methane (at least for the most state points investigated)~\cite{Ranieri2021,KhrapakJMolLiq2022}, and silicon melt at high temperatures~\cite{Luo2022}. Several important non-spherical molecular liquids have been examined using numerical simulations in Ref.~\cite{OhtoriChemLett2020} and  applicability  of the SE relation (\ref{SE_2}) has also been confirmed.    

Water is arguably one of the most ubiquitous and important liquids relevant for our life. Its properties are of great practical significance, but are far from being fully understood. This particularly concerns a number of unusual anomalies exhibited by water in comparison to conventional simple fluids.  One of the important open questions is naturally related to the validity of the SE relation. There is no general consensus regarding which form of the SE relation is particularly suitable and where exactly its breakdown occurs.         

Recently, extensive simulations of the water transport coefficients in a wide range of thermodynamic conditions using the TIP4P/Ice water model have been reported~\cite{BaranJCP2023}. The TIP4P/Ice model was
specifically designed to cope with water near the fluid-solid phase transition and solid-phase properties~\cite{AbascalJCP2005}. In Ref.~\cite{BaranJCP2023} the self-diffusion and shear viscosity coefficients have been evaluated in liquid water in a temperature and pressure ranges from $T=245$ K to $T=350$ K and $P=0$ MPa to $P=500$ MPa, respectively.  
The transport coefficients tabulated in Ref.~\cite{BaranJCP2023} provide an excellent opportunity to examine the validity of the SE relation without the hydrodynamic diameter in the form of Eq.~(\ref{SE_2}).  

\begin{figure}
\includegraphics[width=8.5cm]{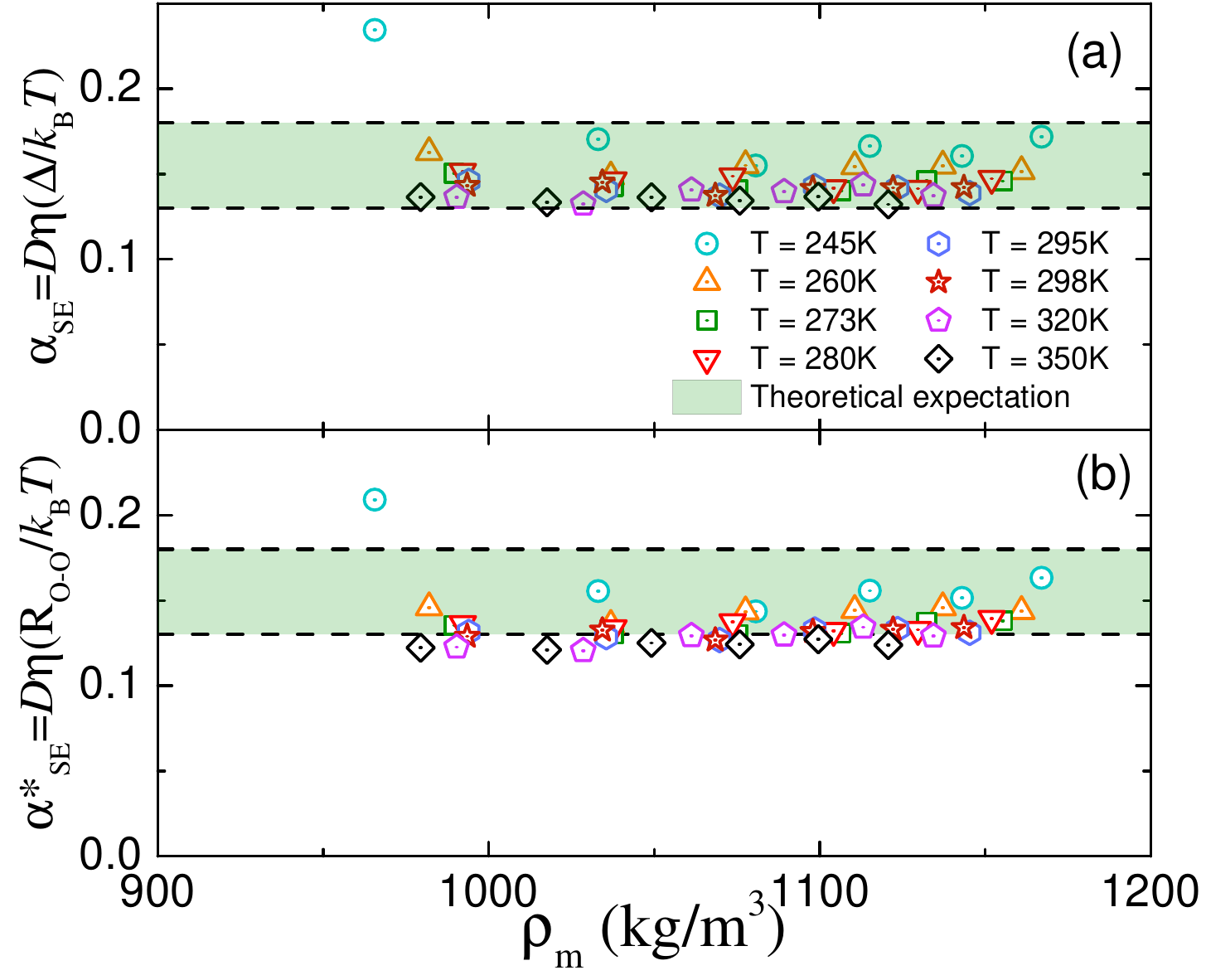}
\caption{(Color online)Stokes-Einstein relation in TIP4P/Ice model of water. (a) The product $D\eta(\Delta/k_{\rm B}T)= \alpha_{\rm SE}$  as a function of mass density $\rho_{\rm m}=m\rho$ along several isotherms investigated in Ref.~\cite{BaranJCP2023}. Shaded region correspond to the theoretically expected range, $0.13\lesssim \alpha_{\rm SE}\lesssim 0.18$. (b) The product $D\eta(R_{\rm O-O}/k_{\rm B}T)$  as a function of mass density along the same isotherms as in (a). Shaded region is the same in (a) and (b) and appears only for comparison in (b). }
\label{Fig1}
\end{figure} 

The results are plotted in Fig~\ref{Fig1}(a), which shows the dependence of $\alpha_{\rm SE}$ on the mass density $\rho_{\rm m}=m\rho$ for eight different temperatures (see label). The (green) shaded region corresponds to the theoretical expectation $0.13<\alpha_{\rm SE}<0.18$. All the data points, except the one at lowest temperature and pressure are confined in the expected region. Namely, we observe that the SE product is constrained to the region $\alpha_{\rm SE}\simeq 0.15\pm 0.02$. Thus, in the parameter regime investigated, the SE relation without the hydrodynamic diameter holds with a good accuracy for the TIP4P/Ice model of water. Remarkably, no free adjustable parameters are involved. Zwanzig's model of SE relation proposed originally for simple fluids applies to a much more complex fluid -- liquid water.  

In Figure~\ref{Fig1}(b) we plot another variant of SE relation, the product $\alpha_{\rm SE}^{*}=D\eta(R_{\rm O-O}/k_{\rm B}T)$, where $R_{\rm O-O}$ is the position of the first maximum of the radial distribution function between oxygen atoms in water. The later is fixed at $R_{\rm O-O}=2.8$ \AA~\cite{BaranJCP2023} and plays the role of hydrodynamic diameter. The (green) shaded region is shown only for comparison with Fig.~\ref{Fig1}(a), because there is no any particular value or range for the quantity $\alpha_{\rm SE}^*$ to be expected a priori. We observe that $\alpha_{\rm SE}^*$ is somewhat smaller than $\alpha_{\rm SE}$ and is constrained to a slightly relatively wider diapason, $\alpha_{\rm SE}^*\simeq 0.14\pm 0.02$. Note that in the mass density range investigated, $970$ kg/m$^3\lesssim \rho_{\rm m}\lesssim 1170$ kg/m$^3$, the characteristic intermolecular separation decreases from $\Delta\simeq 3.1$ \AA~to $\Delta\simeq 2.9$ \AA~only. Thus, taking a fixed length scale of appropriate magnitude (such as $R_{\rm O-O}$ in our example) would not result in a strong violation of the SE relation in the form of Eq.~(\ref{SE_2}). This is, however, expected to happen should wider range of densities be considered.

An important trend observable in Fig.~\ref{Fig1}(a) and (b) is that the SE coefficients slightly but systematically decreases as the temperature increases. The same trend is documented in Fig.~5 of Ref.~\cite{BaranJCP2023}. This might be an indication that the interaction potential somewhat softens with temperature, which leads to lower values of $\alpha_{\rm SE}$, similar to simple fluids (due to lower ratio of shear and longitudinal viscosity and transverse-to-longitudinal sound velocity). This observation also qualitatively correlates with the evidence presented by Ohtori {\it et al}. in Fig. 2 of Ref.~\cite{OhtoriChemLett2020}. They plotted the SE relation for another popular model for water, TIP4P/2005~\cite{AbascalJCP2005/TIP4P2005}, along with selected experimental results. They observed that the SE coefficients remains constant close to $1/2\pi\simeq 0.16$ for $T\lesssim 320$ K and then somewhat drops for higher temperatures. The very fact that the SE relation is valid in the TIP4P/Ice model does not directly follow from its validity in TIP4P/2005 model. The coefficients of self-diffusion and shear viscosity of the two models differ considerably and can only be matched by appropriate rescaling of the temperature, as discussed in Ref.~\cite{BaranJCP2023}. Their product, however, appears model-independent and comply with the SE relation without the hydrodynamic diameter in the investigated parameter regime.

Note that outside the parameter regime studied, the violation of the SE relation upon increasing the temperature and/or lowering density would not be surprising. The SE relation only holds in a limited region of the phase diagram: in sufficiently dense fluids near the fluid-solid phase transition. As an illustration, for the Lennard-Jones (LJ) fluid the SE relation holds for $\rho/\rho_{\rm fr}\gtrsim 0.6$, where $\rho_{\rm fr}$ is the (temperature dependent) freezing density~\cite{KhrapakPRE10_2021} (the onset of validity of SE relation is somewhat higher than the gas-like to liquid-like dynamical crossover also known as Frenkel line on the phase diagram~\cite{BrazhkinPRE2012}, which is roughly located at $\rho/\rho_{\rm fr}\simeq 0.35$ for LJ system~\cite{KhrapakJCP2022}). Another important regime where SE relation is known to be violated concerns strongly supercooled liquids approaching the glass transition, but this situation is beyond the scope of this Note.

To conclude, using recently published data on the self-diffusion and shear viscosity coefficients of bulk liquid water within the TIP4P/Ice model~\cite{BaranJCP2023}, we have demonstrated that the microscopic version of the Stokes-Einstein relation without hydrodynamic diameter is satisfied in the parameter regime investigated. 





\bibliography{SE_Ref}

\providecommand{\noopsort}[1]{}\providecommand{\singleletter}[1]{#1}%
\begin{thebibliography}{28}%
\makeatletter
\providecommand \@ifxundefined [1]{%
 \@ifx{#1\undefined}
}%
\providecommand \@ifnum [1]{%
 \ifnum #1\expandafter \@firstoftwo
 \else \expandafter \@secondoftwo
 \fi
}%
\providecommand \@ifx [1]{%
 \ifx #1\expandafter \@firstoftwo
 \else \expandafter \@secondoftwo
 \fi
}%
\providecommand \natexlab [1]{#1}%
\providecommand \enquote  [1]{``#1''}%
\providecommand \bibnamefont  [1]{#1}%
\providecommand \bibfnamefont [1]{#1}%
\providecommand \citenamefont [1]{#1}%
\providecommand \href@noop [0]{\@secondoftwo}%
\providecommand \href [0]{\begingroup \@sanitize@url \@href}%
\providecommand \@href[1]{\@@startlink{#1}\@@href}%
\providecommand \@@href[1]{\endgroup#1\@@endlink}%
\providecommand \@sanitize@url [0]{\catcode `\\12\catcode `\$12\catcode
  `\&12\catcode `\#12\catcode `\^12\catcode `\_12\catcode `\%12\relax}%
\providecommand \@@startlink[1]{}%
\providecommand \@@endlink[0]{}%
\providecommand \url  [0]{\begingroup\@sanitize@url \@url }%
\providecommand \@url [1]{\endgroup\@href {#1}{\urlprefix }}%
\providecommand \urlprefix  [0]{URL }%
\providecommand \Eprint [0]{\href }%
\providecommand \doibase [0]{http://dx.doi.org/}%
\providecommand \selectlanguage [0]{\@gobble}%
\providecommand \bibinfo  [0]{\@secondoftwo}%
\providecommand \bibfield  [0]{\@secondoftwo}%
\providecommand \translation [1]{[#1]}%
\providecommand \BibitemOpen [0]{}%
\providecommand \bibitemStop [0]{}%
\providecommand \bibitemNoStop [0]{.\EOS\space}%
\providecommand \EOS [0]{\spacefactor3000\relax}%
\providecommand \BibitemShut  [1]{\csname bibitem#1\endcsname}%
\let\auto@bib@innerbib\@empty
\bibitem [{\citenamefont {Balucani}\ and\ \citenamefont
  {Zoppi}(1994)}]{BalucaniBook}%
  \BibitemOpen
  \bibfield  {author} {\bibinfo {author} {\bibfnamefont {U.}~\bibnamefont
  {Balucani}}\ and\ \bibinfo {author} {\bibfnamefont {M.}~\bibnamefont
  {Zoppi}},\ }\href@noop {} {\emph {\bibinfo {title} {Dynamics of the Liquid
  State}}}\ (\bibinfo  {publisher} {Clarendon Press},\ \bibinfo {address}
  {Oxford},\ \bibinfo {year} {1994})\BibitemShut {NoStop}%
\bibitem [{\citenamefont {Costigliola}\ \emph {et~al.}(2019)\citenamefont
  {Costigliola}, \citenamefont {Heyes}, \citenamefont {Schr{\o}der},\ and\
  \citenamefont {Dyre}}]{CostigliolaJCP2019}%
  \BibitemOpen
  \bibfield  {author} {\bibinfo {author} {\bibfnamefont {L.}~\bibnamefont
  {Costigliola}}, \bibinfo {author} {\bibfnamefont {D.~M.}\ \bibnamefont
  {Heyes}}, \bibinfo {author} {\bibfnamefont {T.~B.}\ \bibnamefont
  {Schr{\o}der}}, \ and\ \bibinfo {author} {\bibfnamefont {J.~C.}\ \bibnamefont
  {Dyre}},\ }\bibfield  {title} {\enquote {\bibinfo {title} {Revisiting the
  {S}tokes-{E}instein relation without a hydrodynamic diameter},}\ }\href
  {\doibase 10.1063/1.5080662} {\bibfield  {journal} {\bibinfo  {journal} {J.
  Chem. Phys.}\ }\textbf {\bibinfo {volume} {150}},\ \bibinfo {pages} {021101}
  (\bibinfo {year} {2019})}\BibitemShut {NoStop}%
\bibitem [{\citenamefont {Frenkel}(1955)}]{FrenkelBook}%
  \BibitemOpen
  \bibfield  {author} {\bibinfo {author} {\bibfnamefont {Y.}~\bibnamefont
  {Frenkel}},\ }\href {https://cds.cern.ch/record/106808} {\emph {\bibinfo
  {title} {{Kinetic theory of liquids}}}}\ (\bibinfo  {publisher} {Dover},\
  \bibinfo {address} {New York, NY},\ \bibinfo {year} {1955})\BibitemShut
  {NoStop}%
\bibitem [{\citenamefont {Zwanzig}(1983)}]{ZwanzigJCP1983}%
  \BibitemOpen
  \bibfield  {author} {\bibinfo {author} {\bibfnamefont {R.}~\bibnamefont
  {Zwanzig}},\ }\bibfield  {title} {\enquote {\bibinfo {title} {On the relation
  between self-diffusion and viscosity of liquids},}\ }\href {\doibase
  10.1063/1.446338} {\bibfield  {journal} {\bibinfo  {journal} {J. Chem.
  Phys.}\ }\textbf {\bibinfo {volume} {79}},\ \bibinfo {pages} {4507--4508}
  (\bibinfo {year} {1983})}\BibitemShut {NoStop}%
\bibitem [{\citenamefont {Balucani}\ \emph {et~al.}(1990)\citenamefont
  {Balucani}, \citenamefont {Vallauri},\ and\ \citenamefont
  {Gaskell}}]{Balucani1990}%
  \BibitemOpen
  \bibfield  {author} {\bibinfo {author} {\bibfnamefont {U.}~\bibnamefont
  {Balucani}}, \bibinfo {author} {\bibfnamefont {R.}~\bibnamefont {Vallauri}},
  \ and\ \bibinfo {author} {\bibfnamefont {T.}~\bibnamefont {Gaskell}},\
  }\bibfield  {title} {\enquote {\bibinfo {title} {Generalized
  {S}tokes-{E}instein relation},}\ }\href {\doibase 10.1002/bbpc.19900940313}
  {\bibfield  {journal} {\bibinfo  {journal} {Berichte der Bunsengesellschaft
  f\"{u}r physikalische Chemie}\ }\textbf {\bibinfo {volume} {94}},\ \bibinfo
  {pages} {261--264} (\bibinfo {year} {1990})}\BibitemShut {NoStop}%
\bibitem [{\citenamefont {Khrapak}(2019{\natexlab{a}})}]{KhrapakMolPhys2019}%
  \BibitemOpen
  \bibfield  {author} {\bibinfo {author} {\bibfnamefont {S.}~\bibnamefont
  {Khrapak}},\ }\bibfield  {title} {\enquote {\bibinfo {title}
  {Stokes{\textendash}{E}instein relation in simple fluids revisited},}\ }\href
  {\doibase 10.1080/00268976.2019.1643045} {\bibfield  {journal} {\bibinfo
  {journal} {Mol. Phys.}\ }\textbf {\bibinfo {volume} {118}},\ \bibinfo {pages}
  {e1643045} (\bibinfo {year} {2019}{\natexlab{a}})}\BibitemShut {NoStop}%
\bibitem [{\citenamefont {Khrapak}(2019{\natexlab{b}})}]{KhrapakPRE09_2019}%
  \BibitemOpen
  \bibfield  {author} {\bibinfo {author} {\bibfnamefont {S.}~\bibnamefont
  {Khrapak}},\ }\bibfield  {title} {\enquote {\bibinfo {title} {Elastic
  properties of dense hard-sphere fluids},}\ }\href {\doibase
  10.1103/physreve.100.032138} {\bibfield  {journal} {\bibinfo  {journal}
  {Phys. Rev. E}\ }\textbf {\bibinfo {volume} {100}},\ \bibinfo {pages}
  {032138} (\bibinfo {year} {2019}{\natexlab{b}})}\BibitemShut {NoStop}%
\bibitem [{\citenamefont {Khrapak}\ \emph {et~al.}(2021)\citenamefont
  {Khrapak}, \citenamefont {Kryuchkov}, \citenamefont {Mistryukova},\ and\
  \citenamefont {Yurchenko}}]{KhrapakPRE05_2021}%
  \BibitemOpen
  \bibfield  {author} {\bibinfo {author} {\bibfnamefont {S.}~\bibnamefont
  {Khrapak}}, \bibinfo {author} {\bibfnamefont {N.~P.}\ \bibnamefont
  {Kryuchkov}}, \bibinfo {author} {\bibfnamefont {L.~A.}\ \bibnamefont
  {Mistryukova}}, \ and\ \bibinfo {author} {\bibfnamefont {S.~O.}\ \bibnamefont
  {Yurchenko}},\ }\bibfield  {title} {\enquote {\bibinfo {title} {From soft- to
  hard-sphere fluids: Crossover evidenced by high-frequency elastic moduli},}\
  }\href {\doibase 10.1103/physreve.103.052117} {\bibfield  {journal} {\bibinfo
   {journal} {Phys. Rev. E}\ }\textbf {\bibinfo {volume} {103}},\ \bibinfo
  {pages} {052117} (\bibinfo {year} {2021})}\BibitemShut {NoStop}%
\bibitem [{\citenamefont {Khrapak}\ and\ \citenamefont
  {Khrapak}(2021)}]{KhrapakPRE10_2021}%
  \BibitemOpen
  \bibfield  {author} {\bibinfo {author} {\bibfnamefont {S.~A.}\ \bibnamefont
  {Khrapak}}\ and\ \bibinfo {author} {\bibfnamefont {A.~G.}\ \bibnamefont
  {Khrapak}},\ }\bibfield  {title} {\enquote {\bibinfo {title} {Excess entropy
  and {S}tokes-{E}instein relation in simple fluids},}\ }\href {\doibase
  10.1103/physreve.104.044110} {\bibfield  {journal} {\bibinfo  {journal}
  {Phys. Rev. E}\ }\textbf {\bibinfo {volume} {104}},\ \bibinfo {pages}
  {044110} (\bibinfo {year} {2021})}\BibitemShut {NoStop}%
\bibitem [{\citenamefont {Daligault}(2006)}]{DaligaultPRL2006}%
  \BibitemOpen
  \bibfield  {author} {\bibinfo {author} {\bibfnamefont {J.}~\bibnamefont
  {Daligault}},\ }\bibfield  {title} {\enquote {\bibinfo {title} {Liquid-state
  properties of a one-component plasma},}\ }\href {\doibase
  10.1103/physrevlett.96.065003} {\bibfield  {journal} {\bibinfo  {journal}
  {Phys. Rev. Lett.}\ }\textbf {\bibinfo {volume} {96}},\ \bibinfo {pages}
  {065003} (\bibinfo {year} {2006})}\BibitemShut {NoStop}%
\bibitem [{\citenamefont {Daligault}\ \emph {et~al.}(2014)\citenamefont
  {Daligault}, \citenamefont {Rasmussen},\ and\ \citenamefont
  {Baalrud}}]{DaligaultPRE2014}%
  \BibitemOpen
  \bibfield  {author} {\bibinfo {author} {\bibfnamefont {J.}~\bibnamefont
  {Daligault}}, \bibinfo {author} {\bibfnamefont {K.}~\bibnamefont
  {Rasmussen}}, \ and\ \bibinfo {author} {\bibfnamefont {S.~D.}\ \bibnamefont
  {Baalrud}},\ }\bibfield  {title} {\enquote {\bibinfo {title} {Determination
  of the shear viscosity of the one-component plasma},}\ }\href {\doibase
  10.1103/physreve.90.033105} {\bibfield  {journal} {\bibinfo  {journal} {Phys.
  Rev. E}\ }\textbf {\bibinfo {volume} {90}},\ \bibinfo {pages} {033105}
  (\bibinfo {year} {2014})}\BibitemShut {NoStop}%
\bibitem [{\citenamefont {Khrapak}(2018)}]{KhrapakAIPAdv2018}%
  \BibitemOpen
  \bibfield  {author} {\bibinfo {author} {\bibfnamefont {S.}~\bibnamefont
  {Khrapak}},\ }\bibfield  {title} {\enquote {\bibinfo {title} {Practical
  formula for the shear viscosity of {Y}ukawa fluids},}\ }\href {\doibase
  10.1063/1.5044703} {\bibfield  {journal} {\bibinfo  {journal} {{AIP} Adv.}\
  }\textbf {\bibinfo {volume} {8}},\ \bibinfo {pages} {105226} (\bibinfo {year}
  {2018})}\BibitemShut {NoStop}%
\bibitem [{\citenamefont {Khrapak}(2021)}]{KhrapakMolecules12_2021}%
  \BibitemOpen
  \bibfield  {author} {\bibinfo {author} {\bibfnamefont {S.~A.}\ \bibnamefont
  {Khrapak}},\ }\bibfield  {title} {\enquote {\bibinfo {title} {Self-diffusion
  in simple liquids as a random walk process},}\ }\href {\doibase
  10.3390/molecules26247499} {\bibfield  {journal} {\bibinfo  {journal}
  {Molecules}\ }\textbf {\bibinfo {volume} {26}},\ \bibinfo {pages} {7499}
  (\bibinfo {year} {2021})}\BibitemShut {NoStop}%
\bibitem [{\citenamefont {Heyes}\ and\ \citenamefont
  {Bra{\'{n}}ka}(2007)}]{HeyesPCCP2007}%
  \BibitemOpen
  \bibfield  {author} {\bibinfo {author} {\bibfnamefont {D.~M.}\ \bibnamefont
  {Heyes}}\ and\ \bibinfo {author} {\bibfnamefont {A.~C.}\ \bibnamefont
  {Bra{\'{n}}ka}},\ }\bibfield  {title} {\enquote {\bibinfo {title} {Physical
  properties of soft repulsive particle fluids},}\ }\href {\doibase
  10.1039/b709053f} {\bibfield  {journal} {\bibinfo  {journal} {Phys. Chem.
  Chem. Phys.}\ }\textbf {\bibinfo {volume} {9}},\ \bibinfo {pages} {5570}
  (\bibinfo {year} {2007})}\BibitemShut {NoStop}%
\bibitem [{\citenamefont {Ohtori}\ and\ \citenamefont
  {Ishii}(2015)}]{OhtoriPRE2015}%
  \BibitemOpen
  \bibfield  {author} {\bibinfo {author} {\bibfnamefont {N.}~\bibnamefont
  {Ohtori}}\ and\ \bibinfo {author} {\bibfnamefont {Y.}~\bibnamefont {Ishii}},\
  }\bibfield  {title} {\enquote {\bibinfo {title} {Explicit expression for the
  {S}tokes-{E}instein relation for pure {L}ennard-{J}ones liquids},}\ }\href
  {\doibase 10.1103/physreve.91.012111} {\bibfield  {journal} {\bibinfo
  {journal} {Phys. Rev. E}\ }\textbf {\bibinfo {volume} {91}},\ \bibinfo
  {pages} {012111} (\bibinfo {year} {2015})}\BibitemShut {NoStop}%
\bibitem [{\citenamefont {Ohtori}\ \emph {et~al.}(2017)\citenamefont {Ohtori},
  \citenamefont {Miyamoto},\ and\ \citenamefont {Ishii}}]{OhtoriPRE2017}%
  \BibitemOpen
  \bibfield  {author} {\bibinfo {author} {\bibfnamefont {N.}~\bibnamefont
  {Ohtori}}, \bibinfo {author} {\bibfnamefont {S.}~\bibnamefont {Miyamoto}}, \
  and\ \bibinfo {author} {\bibfnamefont {Y.}~\bibnamefont {Ishii}},\ }\bibfield
   {title} {\enquote {\bibinfo {title} {Breakdown of the {S}tokes-{E}instein
  relation in pure {L}ennard-{J}ones fluids: From gas to liquid via
  supercritical states},}\ }\href {\doibase 10.1103/physreve.95.052122}
  {\bibfield  {journal} {\bibinfo  {journal} {Phys. Rev. E}\ }\textbf {\bibinfo
  {volume} {95}},\ \bibinfo {pages} {052122} (\bibinfo {year}
  {2017})}\BibitemShut {NoStop}%
\bibitem [{\citenamefont {Ohtori}\ \emph {et~al.}(2018)\citenamefont {Ohtori},
  \citenamefont {Uchiyama},\ and\ \citenamefont {Ishii}}]{OhtoriJCP2018}%
  \BibitemOpen
  \bibfield  {author} {\bibinfo {author} {\bibfnamefont {N.}~\bibnamefont
  {Ohtori}}, \bibinfo {author} {\bibfnamefont {H.}~\bibnamefont {Uchiyama}}, \
  and\ \bibinfo {author} {\bibfnamefont {Y.}~\bibnamefont {Ishii}},\ }\bibfield
   {title} {\enquote {\bibinfo {title} {The {S}tokes-{E}instein relation for
  simple fluids: From hard-sphere to {L}ennard-{J}ones via {WCA} potentials},}\
  }\href {\doibase 10.1063/1.5054577} {\bibfield  {journal} {\bibinfo
  {journal} {J. Chem. Phys.}\ }\textbf {\bibinfo {volume} {149}},\ \bibinfo
  {pages} {214501} (\bibinfo {year} {2018})}\BibitemShut {NoStop}%
\bibitem [{\citenamefont {Pieprzyk}\ \emph {et~al.}(2019)\citenamefont
  {Pieprzyk}, \citenamefont {Bannerman}, \citenamefont {Bra{\'{n}}ka},
  \citenamefont {Chudak},\ and\ \citenamefont {Heyes}}]{Pieprzyk2019}%
  \BibitemOpen
  \bibfield  {author} {\bibinfo {author} {\bibfnamefont {S.}~\bibnamefont
  {Pieprzyk}}, \bibinfo {author} {\bibfnamefont {M.~N.}\ \bibnamefont
  {Bannerman}}, \bibinfo {author} {\bibfnamefont {A.~C.}\ \bibnamefont
  {Bra{\'{n}}ka}}, \bibinfo {author} {\bibfnamefont {M.}~\bibnamefont
  {Chudak}}, \ and\ \bibinfo {author} {\bibfnamefont {D.~M.}\ \bibnamefont
  {Heyes}},\ }\bibfield  {title} {\enquote {\bibinfo {title} {Thermodynamic and
  dynamical properties of the hard sphere system revisited by molecular
  dynamics simulation},}\ }\href {\doibase 10.1039/c9cp00903e} {\bibfield
  {journal} {\bibinfo  {journal} {Phys. Chem. Chem. Phys.}\ }\textbf {\bibinfo
  {volume} {21}},\ \bibinfo {pages} {6886--6899} (\bibinfo {year}
  {2019})}\BibitemShut {NoStop}%
\bibitem [{\citenamefont {Li}\ \emph {et~al.}(2021)\citenamefont {Li},
  \citenamefont {Sun}, \citenamefont {Zhang}, \citenamefont {Xian},\ and\
  \citenamefont {Vocadlo}}]{LiJCP2021}%
  \BibitemOpen
  \bibfield  {author} {\bibinfo {author} {\bibfnamefont {Q.}~\bibnamefont
  {Li}}, \bibinfo {author} {\bibfnamefont {T.}~\bibnamefont {Sun}}, \bibinfo
  {author} {\bibfnamefont {Y.}~\bibnamefont {Zhang}}, \bibinfo {author}
  {\bibfnamefont {J.-W.}\ \bibnamefont {Xian}}, \ and\ \bibinfo {author}
  {\bibfnamefont {L.}~\bibnamefont {Vocadlo}},\ }\bibfield  {title} {\enquote
  {\bibinfo {title} {Atomic transport properties of liquid iron at conditions
  of planetary cores},}\ }\href {\doibase 10.1063/5.0062081} {\bibfield
  {journal} {\bibinfo  {journal} {J. Chem. Phys.}\ }\textbf {\bibinfo {volume}
  {155}},\ \bibinfo {pages} {194505} (\bibinfo {year} {2021})}\BibitemShut
  {NoStop}%
\bibitem [{\citenamefont {Ranieri}\ \emph {et~al.}(2021)\citenamefont
  {Ranieri}, \citenamefont {Klotz}, \citenamefont {Gaal}, \citenamefont
  {Koza},\ and\ \citenamefont {Bove}}]{Ranieri2021}%
  \BibitemOpen
  \bibfield  {author} {\bibinfo {author} {\bibfnamefont {U.}~\bibnamefont
  {Ranieri}}, \bibinfo {author} {\bibfnamefont {S.}~\bibnamefont {Klotz}},
  \bibinfo {author} {\bibfnamefont {R.}~\bibnamefont {Gaal}}, \bibinfo {author}
  {\bibfnamefont {M.~M.}\ \bibnamefont {Koza}}, \ and\ \bibinfo {author}
  {\bibfnamefont {L.~E.}\ \bibnamefont {Bove}},\ }\bibfield  {title} {\enquote
  {\bibinfo {title} {Diffusion in dense supercritical methane from
  quasi-elastic neutron scattering measurements},}\ }\href {\doibase
  10.1038/s41467-021-22182-4} {\bibfield  {journal} {\bibinfo  {journal}
  {Nature Commun.}\ }\textbf {\bibinfo {volume} {12}},\ \bibinfo {pages} {1958}
  (\bibinfo {year} {2021})}\BibitemShut {NoStop}%
\bibitem [{\citenamefont {Khrapak}(2022{\natexlab{a}})}]{KhrapakJMolLiq2022}%
  \BibitemOpen
  \bibfield  {author} {\bibinfo {author} {\bibfnamefont {S.A.}\ \bibnamefont
  {Khrapak}},\ }\bibfield  {title} {\enquote {\bibinfo {title} {Diffusion,
  viscosity, and {S}tokes-{E}instein relation in dense supercritical
  methane},}\ }\href {\doibase 10.1016/j.molliq.2022.118840} {\bibfield
  {journal} {\bibinfo  {journal} {J. Mol. Liq.}\ }\textbf {\bibinfo {volume}
  {354}},\ \bibinfo {pages} {118840} (\bibinfo {year}
  {2022}{\natexlab{a}})}\BibitemShut {NoStop}%
\bibitem [{\citenamefont {Luo}\ \emph {et~al.}(2022)\citenamefont {Luo},
  \citenamefont {Zhou}, \citenamefont {Li}, \citenamefont {Lin},\ and\
  \citenamefont {Liu}}]{Luo2022}%
  \BibitemOpen
  \bibfield  {author} {\bibinfo {author} {\bibfnamefont {J.}~\bibnamefont
  {Luo}}, \bibinfo {author} {\bibfnamefont {C.}~\bibnamefont {Zhou}}, \bibinfo
  {author} {\bibfnamefont {Q.}~\bibnamefont {Li}}, \bibinfo {author}
  {\bibfnamefont {Y.}~\bibnamefont {Lin}}, \ and\ \bibinfo {author}
  {\bibfnamefont {L.}~\bibnamefont {Liu}},\ }\bibfield  {title} {\enquote
  {\bibinfo {title} {Atomic transport properties of silicon melt at high
  temperature},}\ }\href {\doibase 10.1016/j.jcrysgro.2022.126701} {\bibfield
  {journal} {\bibinfo  {journal} {J. Crystal Growth}\ }\textbf {\bibinfo
  {volume} {590}},\ \bibinfo {pages} {126701} (\bibinfo {year}
  {2022})}\BibitemShut {NoStop}%
\bibitem [{\citenamefont {Ohtori}\ \emph {et~al.}(2020)\citenamefont {Ohtori},
  \citenamefont {Kondo}, \citenamefont {Shintani}, \citenamefont {Murakami},
  \citenamefont {Nobuta},\ and\ \citenamefont {Ishii}}]{OhtoriChemLett2020}%
  \BibitemOpen
  \bibfield  {author} {\bibinfo {author} {\bibfnamefont {N.}~\bibnamefont
  {Ohtori}}, \bibinfo {author} {\bibfnamefont {Y.}~\bibnamefont {Kondo}},
  \bibinfo {author} {\bibfnamefont {K.}~\bibnamefont {Shintani}}, \bibinfo
  {author} {\bibfnamefont {T.}~\bibnamefont {Murakami}}, \bibinfo {author}
  {\bibfnamefont {T.}~\bibnamefont {Nobuta}}, \ and\ \bibinfo {author}
  {\bibfnamefont {Y.}~\bibnamefont {Ishii}},\ }\bibfield  {title} {\enquote
  {\bibinfo {title} {The {S}tokes-{E}instein relation for non-spherical
  molecular liquids},}\ }\href {\doibase 10.1246/cl.200021} {\bibfield
  {journal} {\bibinfo  {journal} {Chem. Lett.}\ }\textbf {\bibinfo {volume}
  {49}},\ \bibinfo {pages} {379--382} (\bibinfo {year} {2020})}\BibitemShut
  {NoStop}%
\bibitem [{\citenamefont {Baran}\ \emph {et~al.}(2023)\citenamefont {Baran},
  \citenamefont {Rzysko},\ and\ \citenamefont {MacDowell}}]{BaranJCP2023}%
  \BibitemOpen
  \bibfield  {author} {\bibinfo {author} {\bibfnamefont {L.}~\bibnamefont
  {Baran}}, \bibinfo {author} {\bibfnamefont {W.}~\bibnamefont {Rzysko}}, \
  and\ \bibinfo {author} {\bibfnamefont {L.~G.}\ \bibnamefont {MacDowell}},\
  }\bibfield  {title} {\enquote {\bibinfo {title} {Self-diffusion and shear
  viscosity for the {TIP}4{P}/{I}ce water model},}\ }\href {\doibase
  10.1063/5.0134932} {\bibfield  {journal} {\bibinfo  {journal} {J. Chem.
  Phys.}\ }\textbf {\bibinfo {volume} {158}},\ \bibinfo {pages} {064503}
  (\bibinfo {year} {2023})}\BibitemShut {NoStop}%
\bibitem [{\citenamefont {Abascal}\ \emph {et~al.}(2005)\citenamefont
  {Abascal}, \citenamefont {Sanz}, \citenamefont {Fernandez},\ and\
  \citenamefont {Vega}}]{AbascalJCP2005}%
  \BibitemOpen
  \bibfield  {author} {\bibinfo {author} {\bibfnamefont {J.~L.~F.}\
  \bibnamefont {Abascal}}, \bibinfo {author} {\bibfnamefont {E.}~\bibnamefont
  {Sanz}}, \bibinfo {author} {\bibfnamefont {R.~G.}\ \bibnamefont {Fernandez}},
  \ and\ \bibinfo {author} {\bibfnamefont {C.}~\bibnamefont {Vega}},\
  }\bibfield  {title} {\enquote {\bibinfo {title} {A potential model for the
  study of ices and amorphous water: {TIP}4{P}/{I}ce},}\ }\href {\doibase
  10.1063/1.1931662} {\bibfield  {journal} {\bibinfo  {journal} {J. Chem.
  Phys.}\ }\textbf {\bibinfo {volume} {122}},\ \bibinfo {pages} {234511}
  (\bibinfo {year} {2005})}\BibitemShut {NoStop}%
\bibitem [{\citenamefont {Abascal}\ and\ \citenamefont
  {Vega}(2005)}]{AbascalJCP2005/TIP4P2005}%
  \BibitemOpen
  \bibfield  {author} {\bibinfo {author} {\bibfnamefont {J.~L.~F.}\
  \bibnamefont {Abascal}}\ and\ \bibinfo {author} {\bibfnamefont
  {C.}~\bibnamefont {Vega}},\ }\bibfield  {title} {\enquote {\bibinfo {title}
  {A general purpose model for the condensed phases of water:
  {TIP}4{P}/2005},}\ }\href {\doibase 10.1063/1.2121687} {\bibfield  {journal}
  {\bibinfo  {journal} {J. Chem. Phys.}\ }\textbf {\bibinfo {volume} {123}},\
  \bibinfo {pages} {234505} (\bibinfo {year} {2005})}\BibitemShut {NoStop}%
\bibitem [{\citenamefont {Brazhkin}\ \emph {et~al.}(2012)\citenamefont
  {Brazhkin}, \citenamefont {Fomin}, \citenamefont {Lyapin}, \citenamefont
  {Ryzhov},\ and\ \citenamefont {Trachenko}}]{BrazhkinPRE2012}%
  \BibitemOpen
  \bibfield  {author} {\bibinfo {author} {\bibfnamefont {V.~V.}\ \bibnamefont
  {Brazhkin}}, \bibinfo {author} {\bibfnamefont {Yu.~D.}\ \bibnamefont
  {Fomin}}, \bibinfo {author} {\bibfnamefont {A.~G.}\ \bibnamefont {Lyapin}},
  \bibinfo {author} {\bibfnamefont {V.~N.}\ \bibnamefont {Ryzhov}}, \ and\
  \bibinfo {author} {\bibfnamefont {K.}~\bibnamefont {Trachenko}},\ }\bibfield
  {title} {\enquote {\bibinfo {title} {Two liquid states of matter: A dynamic
  line on a phase diagram},}\ }\href {\doibase 10.1103/physreve.85.031203}
  {\bibfield  {journal} {\bibinfo  {journal} {Phys. Rev. E}\ }\textbf {\bibinfo
  {volume} {85}},\ \bibinfo {pages} {031203} (\bibinfo {year}
  {2012})}\BibitemShut {NoStop}%
\bibitem [{\citenamefont {Khrapak}(2022{\natexlab{b}})}]{KhrapakJCP2022}%
  \BibitemOpen
  \bibfield  {author} {\bibinfo {author} {\bibfnamefont {S.~A.}\ \bibnamefont
  {Khrapak}},\ }\bibfield  {title} {\enquote {\bibinfo {title} {Gas-liquid
  crossover in the {L}ennard-{J}ones system},}\ }\href {\doibase
  10.1063/5.0085181} {\bibfield  {journal} {\bibinfo  {journal} {J. Chem.
  Phys.}\ }\textbf {\bibinfo {volume} {156}},\ \bibinfo {pages} {116101}
  (\bibinfo {year} {2022}{\natexlab{b}})}\BibitemShut {NoStop}%
\end{thebibliography}%

\end{document}